\documentclass[12pt]{spieman}  
\usepackage{amsmath,amsfonts,amssymb,mathtools}
\usepackage{graphicx}
\usepackage{setspace}
\usepackage{tocloft}

\title{Vortex coronagraphs for the Habitable Exoplanet Imaging Mission (HabEx) concept: theoretical performance and telescope requirements} 
\usepackage{array}
\newcolumntype{P}[1]{>{\centering\arraybackslash}p{#1}}

\author[a,*]{Garreth Ruane}
\author[a,b]{Dimitri Mawet}
\author[b]{Bertrand Mennesson}
\author[b]{Jeffrey Jewell}
\author[b]{Stuart Shaklan}
\affil[a]{Department of Astronomy, California Institute of Technology, 1200 E. California Blvd., Pasadena, CA 91125, USA}
\affil[b]{Jet Propulsion Laboratory, California Institute of Technology, 4800 Oak Grove Dr., Pasadena, CA 91109, USA}

\cftpagenumbersoff{figure}
\cftpagenumbersoff{table}

\graphicspath{{Figures/}}
 
\begin{document} 
  \maketitle 

\begin{abstract}
The Habitable Exoplanet Imaging Mission (HabEx) concept requires an optical coronagraph that provides deep starlight suppression over a broad spectral bandwidth, high throughput for point sources at small angular separation, and insensitivity to temporally-varying, low-order aberrations. Vortex coronagraphs are a promising solution that perform optimally on off-axis, monolithic telescopes and may also be designed for segmented telescopes with minor losses in performance. We describe the key advantages of vortex coronagraphs on off-axis telescopes: 1) Unwanted diffraction due to aberrations is passively rejected in several low-order Zernike modes relaxing the wavefront stability requirements for imaging Earth-like planets from $<$10 to $>$100~pm~rms. 2) Stars with angular diameters $>$0.1~$\lambda/D$ may be sufficiently suppressed. 3) The absolute planet throughput is $>$10\%, even for unfavorable telescope architectures. 4) Broadband solutions ($\Delta\lambda/\lambda>0.1$) are readily available for both monolithic and segmented apertures. The latter make use of grayscale apodizers in an upstream pupil plane to provide suppression of diffracted light from amplitude discontinuities in the telescope pupil without inducing additional stroke on the deformable mirrors. We set wavefront stability requirements on the telescope, based on a stellar irradiance threshold set at an angular separation of $3\pm0.5\lambda/D$ from the star, and discuss how some requirements may be relaxed by trading robustness to aberrations for planet throughput. 
\end{abstract}


{\noindent \footnotesize\textbf{*}NSF Astronomy and Astrophysics Postdoctoral Fellow, \linkable{gruane@caltech.edu}}


\section{Introduction}
\label{sec:intro}  

The Habitable Exoplanet Imaging Mission (HabEx) concept seeks to directly detect atmospheric biomarkers on Earth-like exoplanets orbiting sun-like stars for the first time\cite{Mennesson2016}. Accomplishing this task requires extremely high-contrast imaging over a broad spectral range using an internal coronagraph \cite{exoc} or external starshade \cite{exos}. Sufficient starlight suppression may be achieved on an ultra-stable telescope using an on-board coronagraph instrument with high-precision wavefront control and masks specially designed to manage diffraction of unwanted starlight. Each of these critical technologies will be demonstrated in space with the WFIRST coronagraph instrument (CGI) at the performance level needed to image gas giant planets in reflected light with a 2.4~m telescope \cite{wfirst}. Leveraging the advancements afforded by WFIRST, the HabEx mission concept makes use of a larger ($>$4~m) telescope whose stability specifications allow for the detection and characterization Earth-like planets with planet-to-star flux ratios $<10^{-10}$.

The optimal coronagraph performance for a given mission depends strongly on the telescope design. The possible HabEx architectures currently under study in preparation for the 2020 Astrophysics Decadal Survey are a 4~m monolithic (architecture A) or 6.5~m segmented primary mirror (architecture B). For the purposes of this paper, we assume fully off-axis telescopes in both cases. The unobstructed, circular pupil provided by architecture A is ideal for coronagraph performance. On the other hand, the segmented primary mirror of architecture B will introduce a number of additional complications owing to potential amplitude and phase discontinuities in the wavefront. 

We present vortex coronagraph\cite{Mawet2005,Foo2005,Mawet2010a} designs for each HabEx architecture and their theoretical performance. We demonstrate that the coronagraph may be designed to passively reject unwanted diffraction within the telescope in the presence of temporally-varying, low-order aberrations as well as amplitude discontinuities (i.e. gaps between mirror segments). We set wavefront stability requirements on the telescope, including the phasing of the primary mirror segments in the case of architecture B, and discuss how some telescope requirements may be relaxed by trading robustness to aberrations for planet throughput \cite{Green2003}. 

\section{Architecture A: 4~m off-axis, unobscured, monolithic telescope} \label{sec:monolith} 

\begin{figure}[t]
    \centering
    \includegraphics[width=\linewidth]{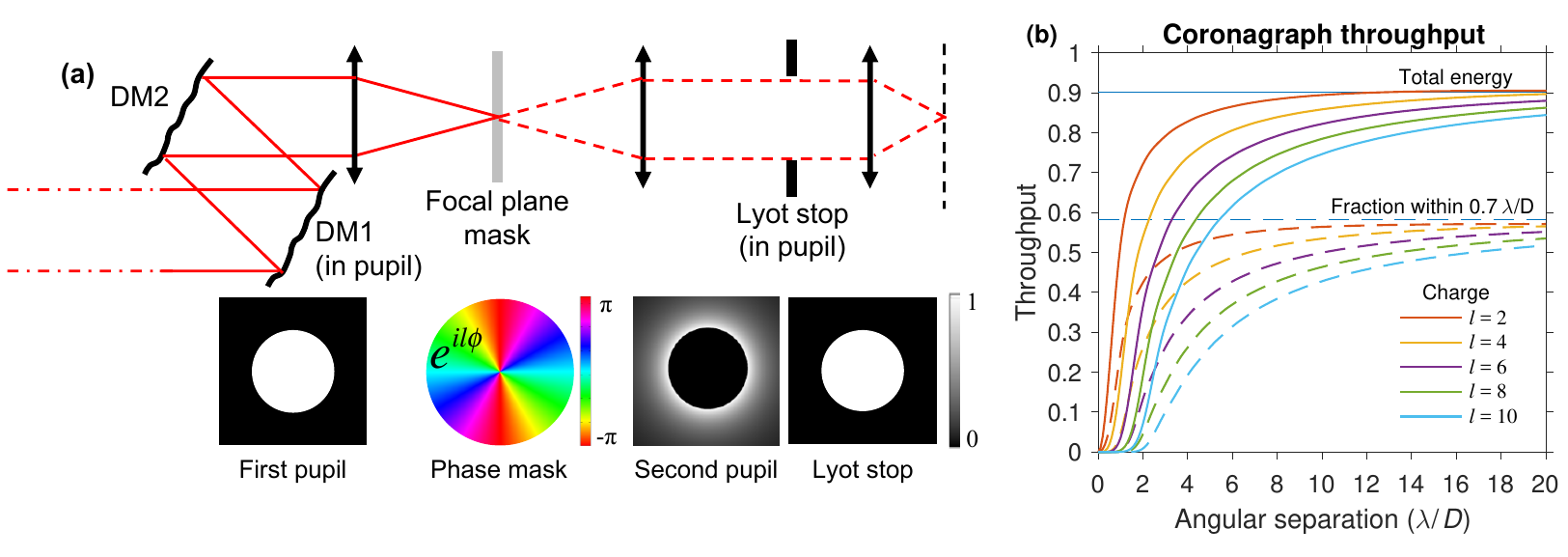}
    \caption{(a) Schematic of a vortex coronagraph with deformable mirrors DM1 and DM2, focal plane phase mask with complex transmittance $\exp(il\phi)$, and circular Lyot stop. Starlight suppression is achieved by diffracting the stellar field outside of the Lyot stop. (b) Throughput performance of a vortex coronagraph for Lyot stop whose radius is 95\% that of the geometric pupil ($b/a=0.95$). The horizontal lines indicate the maximum throughput in each case.}
    \label{fig:VCschematic}
\end{figure}

The first telescope architecture we analyze is a 4~m off-axis telescope with a monolithic primary mirror. The unobstructed pupil is conducive to highly efficient coronagraph designs, such as the vortex coronagraph, which provide sensitivity to weak planet signals at small angular separations, as demonstrated in the laboratory \cite{Mawet2009} and observations with ground based telescopes \cite{Mawet2010b,Serabyn2010,Serabyn2017,Mawet2017,Ruane2017}. Figure \ref{fig:VCschematic}a shows a schematic of a vortex coronagraph with dual deformable mirrors for wavefront control, a focal plane mask, and Lyot stop. The vortex focal plane mask is a transparent optic which imparts a spiral phase shift of the form $\exp(il\phi)$ on the incident field, where $l$ is an even non-zero integer known as the ``charge" and $\phi$ is the azimuth angle in the focal plane. Light from an on-axis point source (i.e. the star) that passes through the circular entrance pupil of radius $a$ is completely diffracted outside of the downstream Lyot stop of radius $b$, assuming $b<a$ and one-to-one magnification within the coronagraph. In addition to ideal starlight suppression, the vortex coronagraph provides high throughput for point-like sources at small angular separations from the star (see Fig. \ref{fig:VCschematic}b).

\subsection{Ideal coronagraph throughput}


We present two common throughput definitions in the literature: (1)~the fraction of planet energy from a planet that reaches the image plane and (2)~the fraction of the planet energy that falls within a circular region-of-interest with radius $\hat{r}\lambda/D$ centered at the planet position, where $\lambda$ is the wavelength and $D$ is the diameter of the primary mirror. The maximum throughput (at large angular separations) in each case is
\begin{equation}
    \eta_{p,\text{max}}=
    \begin{cases}
        (b/a)^2, & \text{total energy}\\
        (b/a)^2\left[1-J_0\left(\pi \hat{r}  \frac{b}{a}\right)^2-J_1\left(\pi \hat{r} \frac{b}{a}\right)^2\right], & \text{fraction within } \hat{r} \lambda/D \text{ radius}\\
    \end{cases},
\end{equation}
where $J_0(~)$ and $J_1(~)$ are Bessel functions of the first kind\cite{bornwolf}. For example, if $b/a=0.95$ and $\hat{r}=0.7$, the theoretical maxima for case (1) and (2) are 90\% and 58\%, respectively. The latter value may also be normalized to the same quantity without the coronagraph masks. For example, in the case described above, 86\% of planet energy remains within 0.7~$\lambda/D$ of the planet's position in the image, a value referred to as the relative throughput. In the remainder of this section, we assume a typical value of $b/a=0.95$ for architecture A. In practice, the value of $b/a$ will be selected based on the desired tolerance to lateral pupil motion and magnification. Definitions (1) and (2) are plotted for various values of $l$ in Fig. \ref{fig:VCschematic}b for angular separations up to 20~$\lambda/D$ using numerical beam propagation.

\subsection{Passive insensitivity to low-order aberrations}

Detecting Earth-like exoplanets in practice will require a coronagraph whose performance is insensitive to wavefront errors owing to mechanical motions in the telescope and differential polarization aberrations, which both manifest as low-order wavefront errors. We describe the phase at the entrance pupil of the coronagraph as a linear combination of Zernike polynomials $Z_n^m (r/a,\theta)$ defined over a circular pupil of radius $a$. An isolated phase aberration is written
\begin{equation}
P(r,\theta) = \exp \left[i c_{nm} Z_n^m(r/a,\theta)\right],\;\;\;\;\;r\le a,
\end{equation}
where $i=\sqrt{-1}$ and $c_{n,m}$ is the Zernike coefficient. Assuming small wavefront errors (i.e. $c_{nm}\ll$~1~rad~rms), the field in the pupil may be approximated to first order via its Taylor series expansion:
\begin{equation}
P(r,\theta) \approx 1 + i c_{nm} Z_n^m(r/a,\theta),\;\;\;\;\;r\le a.
\label{eq:Ztaylor}
\end{equation}
For convenience, we choose to use the set real-valued of Zernike polynomials described by
\begin{equation}
    Z_n^m(r/a,\theta) = R_n^{|m|}(r/a)q_m(\theta)
    ,\;\;\;\;\;r\le a,
\label{eq:Zpupil}
\end{equation}
where $R_n^m\left(r/a\right)$ are the radial polynomials described in Appendix A and 
\begin{equation}
    q_m(\theta) = 
    \begin{cases} 
        \cos(m\theta) & m \geq 0 \\
        \sin(|m|\theta) & m<0
    \end{cases}
    .
\label{eq:q}
\end{equation}
The field transmitted through a vortex phase element of charge $l$, owing to an on-axis point source, is given by the product of $\exp\left(il\phi\right)$ and the optical Fourier transform (FT) of Eq. \ref{eq:Ztaylor}:
\begin{equation}
F_{nml}(\rho,\phi)\approx\left[f_{00}(\rho,\phi) + ic_{nm}f_{nm}(\rho,\phi) \right]e^{il\phi},
\label{eq:ZPSF}
\end{equation}
where
\begin{equation}
f_{nm}(\rho,\phi) =\frac{k a^2}{f}\frac{J_{n+1}\left( k a \rho/f\right)}{k a \rho/f} q_m(\phi),
\label{eq:ZPSF2}
\end{equation}
$\rho$ is the radial polar coordinate in the focal plane, $k=2\pi/\lambda$, $\lambda$ is the wavelength, and $f$ is the focal length. The field in the subsequent pupil plane (i.e. just before the Lyot stop), $E_{lnm}$, is given by the FT of Eq. \ref{eq:ZPSF}. The first term, $f_{00}(\rho,\phi)$, is the common Airy pattern, which diffracts completely outside of the Lyot stop for all even nonzero values of $l$. In this case, the Lyot plane field becomes\cite{Carlotti2009}
\begin{equation}
E_{l,\mathrm{Airy}}(r,\theta)=
    \begin{cases} 
        0 & r \le a \\
        \frac{a}{r}R_{|l|-1}^1(\frac{a}{r})e^{il\theta} & r>a
    \end{cases}.
\label{eqn:El00}
\end{equation}
More generally, the full Lyot plane field is given by
\begin{equation}
E_{nml}(r,\theta)\approx E_{l,\mathrm{Airy}}(r,\theta) + ic_{nm}g_{nml}(r,\theta),
\label{eqn:Elnm}
\end{equation}
where
\begin{equation}
g_{nml}(r,\theta)=\frac{k a}{2 f}e^{il\theta}
    \begin{cases} 
        (-1)^m e^{im\theta}\mathcal{W}_{n+1}^{l+m}(r)+e^{-im\theta}\mathcal{W}_{n+1}^{l-m}(r) & m \ge 0 \\
        i\left[(-1)^{m+1} e^{im\theta}\mathcal{W}_{n+1}^{l+m}(r)+e^{-im\theta}\mathcal{W}_{n+1}^{l-m}(r)\right] & m<0
    \end{cases},
\label{eqn:glnm}
\end{equation}
and $\mathcal{W}_p^q(r)$ is a special case of the Weber-Schafheitlin integral (Appendix B):
\begin{equation}
\begin{split}
\mathcal{W}_p^q(r)=&W_{p,q,0}(r;{k a \rho}/{f},{k r \rho}/{f})\\
=&\int\limits_{0}^{\infty }{J_p\left({k a \rho}/{f}\right)J_q\left( {k r \rho}/{f}\right)d\rho}.
\end{split}
\end{equation}
\begin{figure}[t!]
    \centering
    \includegraphics[width=3.1in]{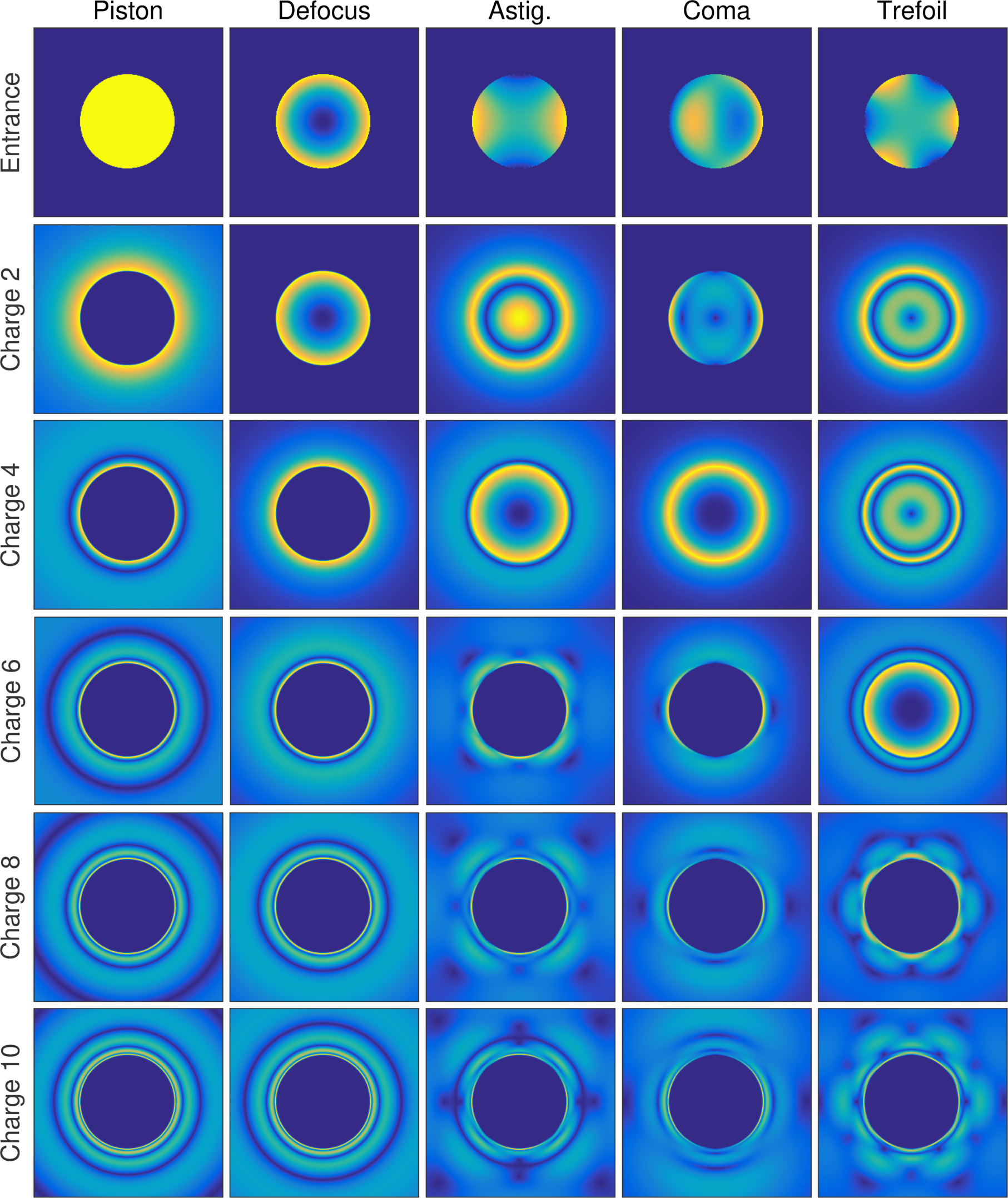}
    \caption{The low-order aberration filtering mechanism of a vortex coronagraph. The top row shows the wavefront at the entrance pupil of the coronagraph (``first pupil" in Fig. \ref{fig:VCschematic}). The remaining rows show the amplitude distribution just before the Lyot stop (``second pupil" in Fig. \ref{fig:VCschematic}). A vortex coronagraph is passively insensitive to modes where the starlight appears outside of a Lyot stop whose radius $b$ is less than the geometric pupil radius $a$ and $|l|>n+|m|$.}
    \label{fig:Zfig}
\end{figure}
The solutions to Eq. \ref{eqn:glnm} are shown in Fig. \ref{fig:Zfig} and listed in Appendix C. In cases where all of the light is located outside of the geometric pupil, the source is extinguished by a Lyot stop with radius $b<a$. The constant term in Eq. \ref{eq:Ztaylor} is completely suppressed for all nonzero even values of $l$. However, the first order term is also blocked by the Lyot stop if $|l|>n+|m|$. A charge $l$ vortex coronagraph is therefore passively insensitive to the $l^2/4$ Zernike modes rejected by the Lyot stop.

\subsection{Wavefront stability requirements}

The wavefront error tolerances of a given coronagraph design depend on the aberration mode and/or spatial frequency content of the error. The coronagraph and telescope must be jointly optimized to passively suppress starlight and provide the stability needed to maintain suppression throughout an observation. We present telescope stability requirements for Earth-like exoplanet imaging with vortex coronagraphs in terms of low-order and mid-to-high spatial frequency aberrations.

\begin{figure}[t!]
    \centering
    \includegraphics[scale=0.7]{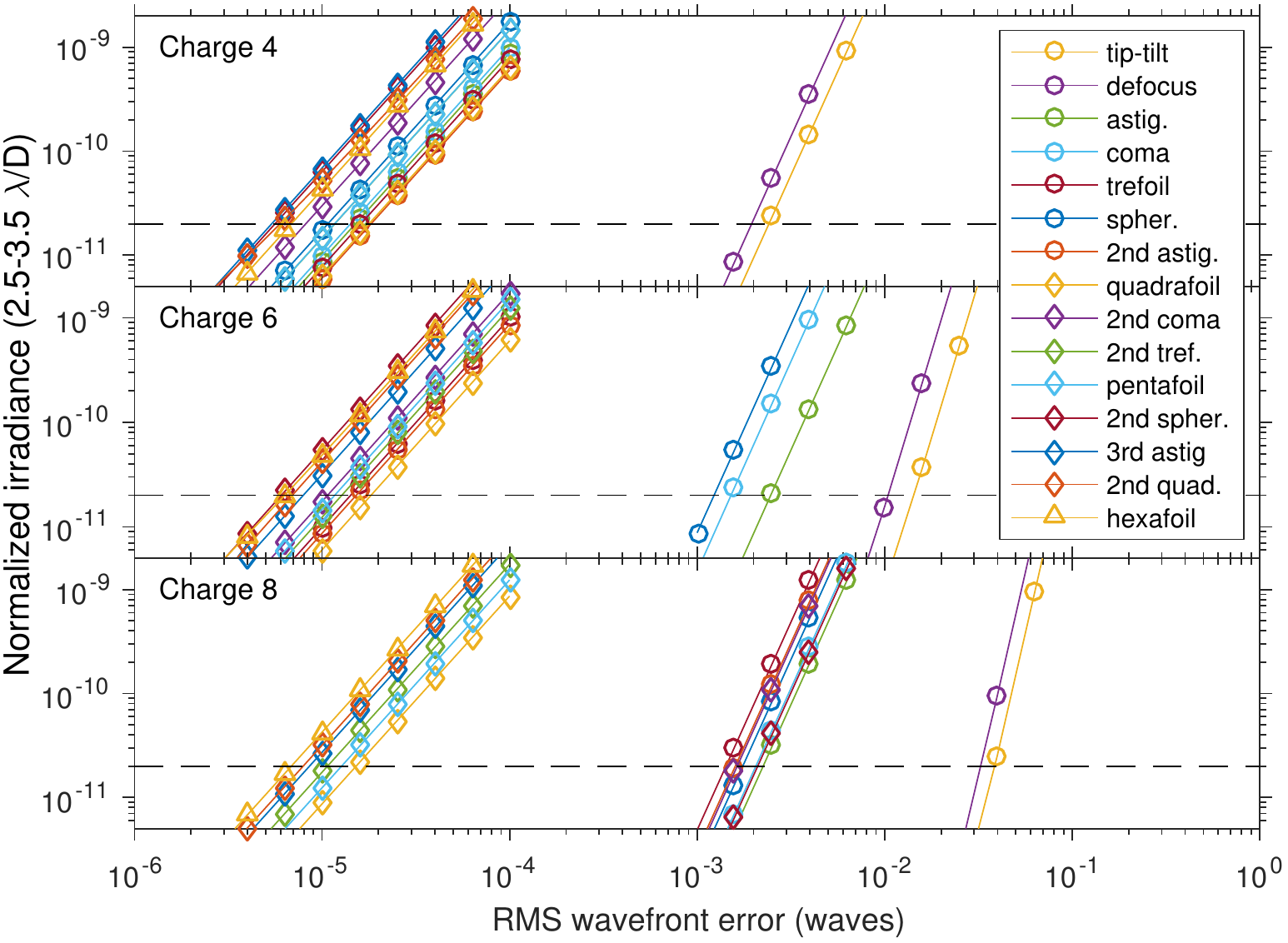}
    \caption{Sensitivity of vortex coronagraphs to low-order aberrations. The stellar irradiance is averaged over effective angular separations $3\pm0.5~\lambda/D$, normalized to the peak irradiance without the coronagraph masks, as a function of root-mean-square (RMS) wavefront error in each Zernike aberration. As the vortex charge increases, larger errors may be tolerated on the lowest order aberrations, which typically dominate the dynamic wavefront error budget.}
    \label{fig:IrrVsWFE}
\end{figure}

\subsubsection{Low-order requirements: Zernike aberrations}

Figure \ref{fig:IrrVsWFE} shows the leaked starlight through the coronagraph (stellar irradiance, averaged over effective angular separations $3\pm0.5~\lambda/D$, and normalized to the peak value without the coronagraph masks) as a function of root-mean-square (rms) wavefront error. Modes with $n+|m|\ge |l|$ follow a quadratic power law and generate irradiance at the $\sim10^{-11}$ level for wavefront errors of $\sim10^{-5}$ waves rms. However, modes with $n+|m|<|l|$ are blocked at least to first order at the Lyot stop, as described in the previous section. In these cases, the equivalent irradiance level ($\sim10^{-11}$) corresponds to $\sim100\times$ the wavefront error. 

\begin{figure}[t!]
    \centering
    \includegraphics[width=0.8\linewidth]{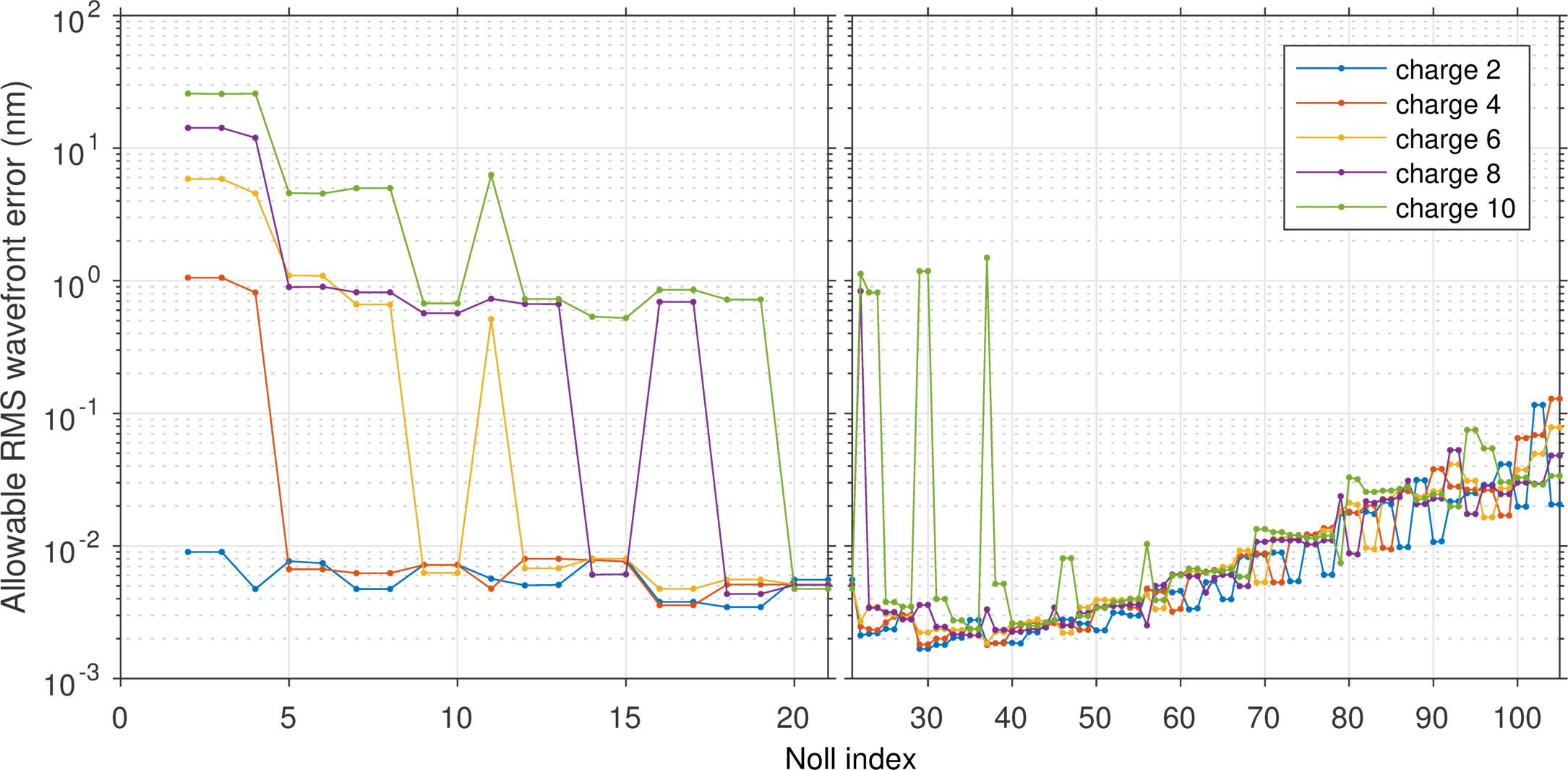}
    \caption{Wavefront error requirements in the Zernike mode basis. The maximum allowable RMS wavefront error generates a normalized irradiance of 2$\times10^{-11}$ at an effective separation of $3\pm0.5~\lambda/D$ for $\lambda=450$~nm. The Zernike modes are ordered by Noll index\cite{Noll1976} to conform to conventions in astronomy. However, the wavefront error tolerance depends more naturally on the sum of indices $n+|m|$. As the charge increases, large wavefront errors ($>$100~pm~rms) may be tolerated on more of the low order aberrations. We have emphasized the lowest 21 modes, many of which tend to dominate the wavefront error budget and may be readily suppressed by vortex coronagraphs. The requirements for Noll indices $>$40 are roughly the same for all charges $\le 10$.}
    \label{fig:WFEvsNoll}
\end{figure}

We place requirements on the stability of the wavefront by setting a maximum allowable irradiance threshold on the leaked starlight at $3\pm0.5~\lambda/D$. This angular coordinate range corresponds to the separations where a charge~8 vortex coronagraph transmits $\sim$50\% of the planet light. Here, the threshold is chosen to be 2$\times10^{-11}$ per Zernike mode (dashed line in Fig. \ref{fig:IrrVsWFE}) to prevent any single low-order aberration from dominating the error budget. The corresponding wavefront error at $\lambda=450$~nm, likely the shortest and most challenging wavelength, are shown in Fig. \ref{fig:WFEvsNoll} and listed in Table \ref{tab:loworder}. Modes that are passively suppressed by the coronagraph have wavefront requirements $>100$~pm~rms, while those that transmit tend to require $<$10~pm~rms. The minimum charge of the vortex coronagraph may be chosen to preserve robustness to particularly problematic low-order aberrations as well as to relax requirements and reduce the cost of the overall mission. However, increasing the minimum charge has a significant impact on the scientific yield of the mission, especially since insufficient throughput at small angular separations (i.e. beyond the so-called ``inner working angle") will likely limit the number of detected and characterized Earth-like planets within the mission lifetime \cite{Stark2015}.

The requirements given in Fig. \ref{fig:WFEvsNoll} and Table \ref{tab:loworder} may be scaled to any wavelength by simply multiplying the reported rms wavefront error by a factor of $\lambda/(450~\text{nm})$. While a higher charge (e.g. charge 6 or 8) may be used for the shortest wavelengths to improve robustness, using a lower charge (e.g. charge 4) at longer wavelengths would allow exoplanets detected near the inner working angle of the visible coronagraph to be characterized in the infrared, where the wavefront error requirements are naturally less strict. In that case, the infrared coronagraph would drive requirements in some of the lowest order modes, which would be relaxed by a factor of $\gtrsim$2 with respect to higher-order requirements driven by the visible coronagraph.  

\subsubsection{Mid-to-high spatial frequency requirements}

\begin{figure}[t!]
    \centering
    \includegraphics[scale=0.6]{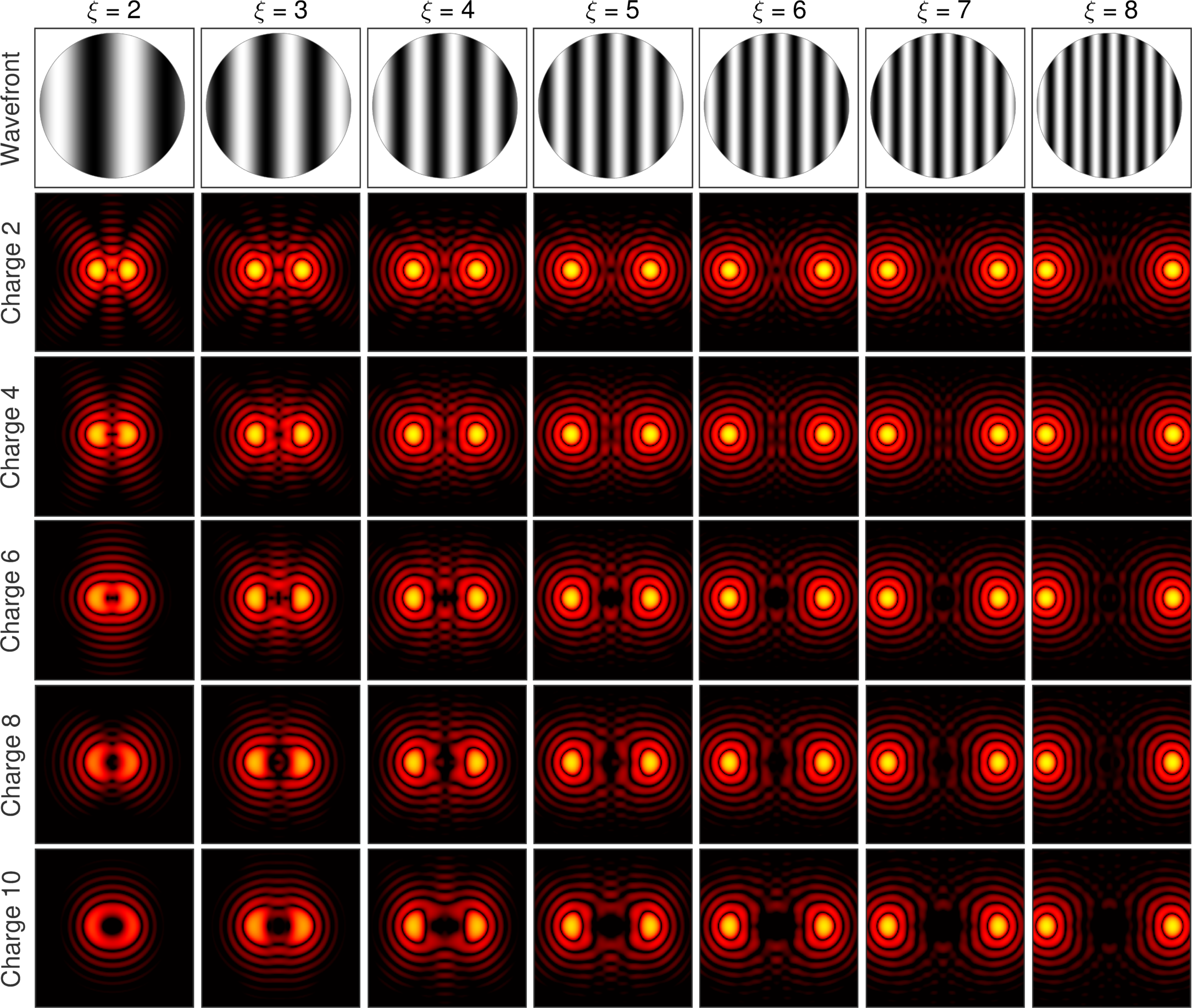}
    \caption{Stellar irradiance (log scale) due to a sinusoidal phase error as a function of spatial frequency, $\xi$, and charge. An error in a single Fourier mode generates a speckle at the corresponding position in the image plane. For example, if $\omega\lambda=100~\text{pm}$, the raw contrast at $x^\prime=\pm\xi\lambda F^\#$ is $C \approx 2 (\pi \omega)^2=9.7\times10^{-7}$ at $\lambda=450~\text{nm}$ regardless of the charge.}
    \label{fig:SpatFreqLeakage_grid}
\end{figure}

Whereas the coronagraph design provides degrees of freedom for controlling robustness to low-order aberrations, high throughput coronagraphs are naturally sensitive to mid- and high-spatial frequency aberrations. In fact, any coronagraph which passively suppresses mid-spatial frequency aberrations must also have low throughput for off-axis planets. This is an outcome of the well known relationship between raw contrast and the RMS wavefront error in Fourier modes\cite{Malbet1995}. The pupil field associated with a single spatial frequency is given by
\begin{align}
P(r,\theta) & = \exp \left[i 2\sqrt{2}\pi \omega \sin\left(\frac{2 \pi x}{a} \xi  \right)\right],\;\;\;\;\;r\le a, \\
& \approx 1 + i 2\sqrt{2}\pi \omega \sin\left(\frac{2 \pi x}{a} \xi\right),\;\;\;\;\;r\le a,
\end{align}
where $r^2=x^2+y^2$, $\xi$ is the spatial frequency in cycles per pupil diameter, and $\omega$ is the RMS phase error in waves where we have assumed $\omega \ll 1$. The corresponding field just before the focal plane mask is
\begin{equation}
\begin{split}
F(\rho,\phi) &= f_{00}(\rho,\phi)\\
&+\sqrt{2} \pi \omega \left[ f_{00}(\rho_-,\phi) - f_{00}(\rho_+,\phi) \right],
\end{split}
\end{equation}
where $\rho^2=x^{\prime2}+y^{\prime2}$, $\rho_-^2=(x^\prime-\xi\lambda F^\#)^2+y^2$, $\rho_+^2=(x^\prime+\xi\lambda F^\#)^2+y^2$, and $F^\# = f/(2a)$. The coronagraph completely rejects the $f_{00}(\rho,\phi)$ term. Thus, at position $(x^\prime,y^\prime)=(\xi \lambda F^\#,0)$ after the coronagraph
\begin{equation}
F(\xi\lambda F^\#,0) = \sqrt{2\eta_p} \pi \omega \left[ f_{00}\left(0,0\right) - f_{00}\left(2\xi\lambda F^\#,0\right)  \right],
\end{equation}
where $\eta_p$ is the coronagraph throughput and $F^\# = f/(2b)$. Solving for the normalized stellar irradiance, $\eta_s$, we find
\begin{equation}
\eta_s = \eta_p 2 (\pi \omega)^2 \frac{\left| f_{00}\left(0,0\right) - f_{00}\left(2\xi\lambda F^\#,0\right)  \right|^2}{\left| f_{00}\left(0,0\right) \right|^2}.
\end{equation}
Therefore, for $\xi \gtrsim 1$, the raw contrast at $(x^\prime,y^\prime)=(\xi \lambda F^\#,0)$ is
\begin{equation}
C = \eta_s/\eta_p \approx 2 (\pi \omega)^2.
\end{equation}
For example, a 1 pm rms mid-spatial frequency wavefront error described by the vector $\vec{\xi}=\xi_x \hat{x} + \xi_y \hat{y}$ generates a change in raw contrast of $\sim10^{-10}$ at $\lambda=450~\text{nm}$ in the corresponding image plane location $(x^\prime,y^\prime)=(\xi_x \lambda F^\# ,\xi_y \lambda F^\#)$. This implies a stability requirement of $\sim$1~pm~rms per Fourier mode for mid-spatial frequency wavefront errors. The stellar irradiance as a function of spatial frequency and charge is shown in Fig. \ref{fig:SpatFreqLeakage_grid}.

Rejecting starlight with mid-spatial frequency phase errors and proportionally reducing the coronagraph throughput at the position of interest degrades performance in the photon-noise-limited regime where the signal-to-noise ratio (SNR) for planet detection scales as $\eta_p/\sqrt{\eta_s}$. Provided an optimal coronagraph maximizes the SNR, a coronagraph that is passively robust to spatial frequencies where $(x^\prime,y^\prime)=(\xi_x \lambda F^\# ,\xi_y \lambda F^\#)$ is in the region of interest (i.e. dark hole) is not desirable.

\subsection{Sensitivity to partially resolved, extended sources}

\begin{figure}
    \centering
    \includegraphics[width=\linewidth]{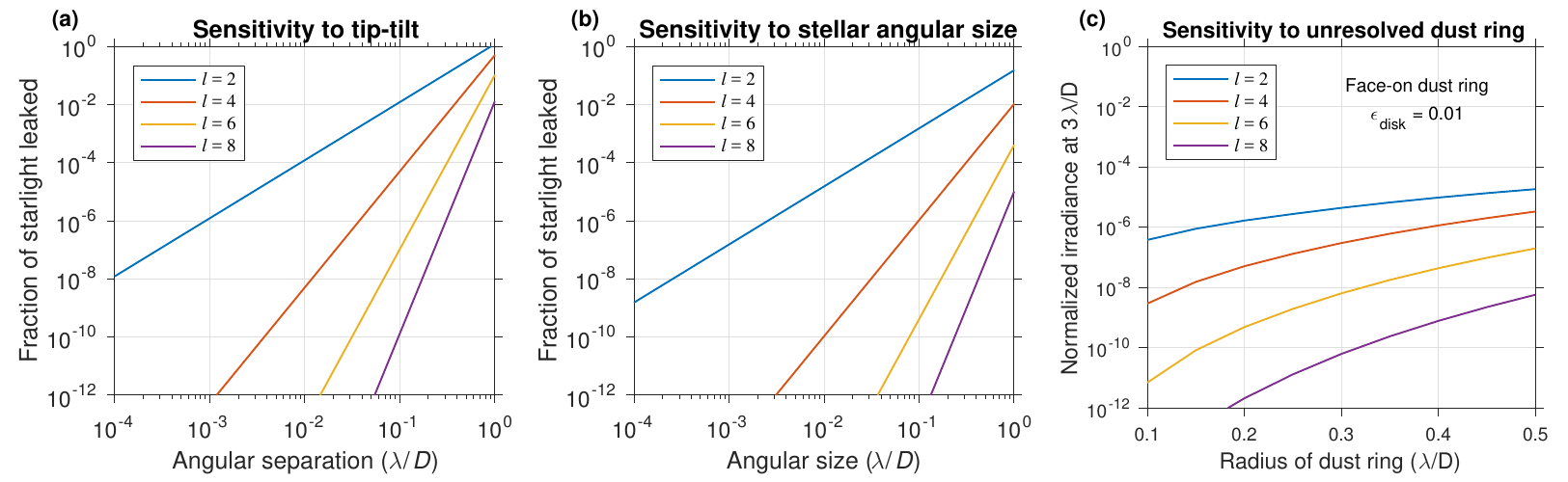}
    \caption{Sensitivity of the vortex coronagraph to (a) tip-tilt, (b) stellar angular size, and (c) unresolved dust rings. (a)-(b)~The total energy leaked versus (a) the angular separation of a point source and (b) the angular size of the star. The coefficients for each power law are given in Table \ref{tab:tiptilt_coeffs}. (c) Stellar irradiance, averaged over source positions $3\pm0.5~\lambda/D$ and normalized to the peak of the telescope PSF, owing to an unresolved ring of dust at astrophysical contrast of $\epsilon=1\%$.}
    \label{fig:leakfromextendedsources}
\end{figure}

The fraction of energy from a point source that leaks through the coronagraph as a function of angular separation, $\alpha$, may be approximated for small offsets (i.e. $\alpha \ll \lambda/D$) through modal decomposition of the source \cite{Ruane2016dissertation}. The transmitted energy is given by $T_{\alpha}=\tau_l (\pi \alpha D/\lambda)^l$, where $\tau_l$ is a constant (see Fig. \ref{fig:leakfromextendedsources}a). Integrating over an extended, spatially incoherent, stellar source of angular extent, $\Theta$, the expression becomes $T_{\Theta}=\kappa_l (\pi \Theta D/\lambda)^l$, where $\kappa_l$ is a constant (see Fig. \ref{fig:leakfromextendedsources}b). The theoretical values for $\tau_l$ and $\kappa_l$ are given in Table \ref{tab:tiptilt_coeffs}. Higher charge vortex coronagraphs are far less sensitive to small tip-tilt errors and sources of finite size. For example, charge 6 vortex coronagraphs sufficiently suppress light from stars with angular diameters up to $\sim$0.1~$\lambda/D$ or $\sim$2~mas for a 4~m telescope at $\lambda=450$~nm.

An often overlooked potential source of leaked starlight is the presence of an unresolved disk of dust around the star. Figure \ref{fig:leakfromextendedsources}c shows the stellar irradiance that appears in the image plane due to scattered light from the debris ring at astrophysical contrast of 1\% and angular separation of 3~$\lambda/D$ from the star as a function of the size of the ring. For example, imaging an Earth-like planet at 3~$\lambda/D$ around a star with a dust ring of radius $0.2~\lambda/D$ requires at least a charge 6 coronagraph. We note that over the last 10 years, long baseline near infrared interferometric observations \cite{Absil2013,Ertel2014} have suggested that $\sim$10-20\% of nearby main sequence stars have such rings of small hot dust grains concentrating near the sublimation radius, and contributing about 1\% off the total solar flux in the near infrared. Assuming that we are looking for a 300~K planet at $3~\lambda/D$ separation, 1500K dust grains would be located 25$\times$ closer; i.e at $0.12~\lambda/D$. In addition to being more resilient to low-order wavefront aberrations, higher charge vortex coronagraphs are also less sensitive to astrophysical noise sources in the inner part of the system, such as bright dust rings that may be fairly common. 

\begin{figure}
    \centering
    \includegraphics[width=\linewidth]{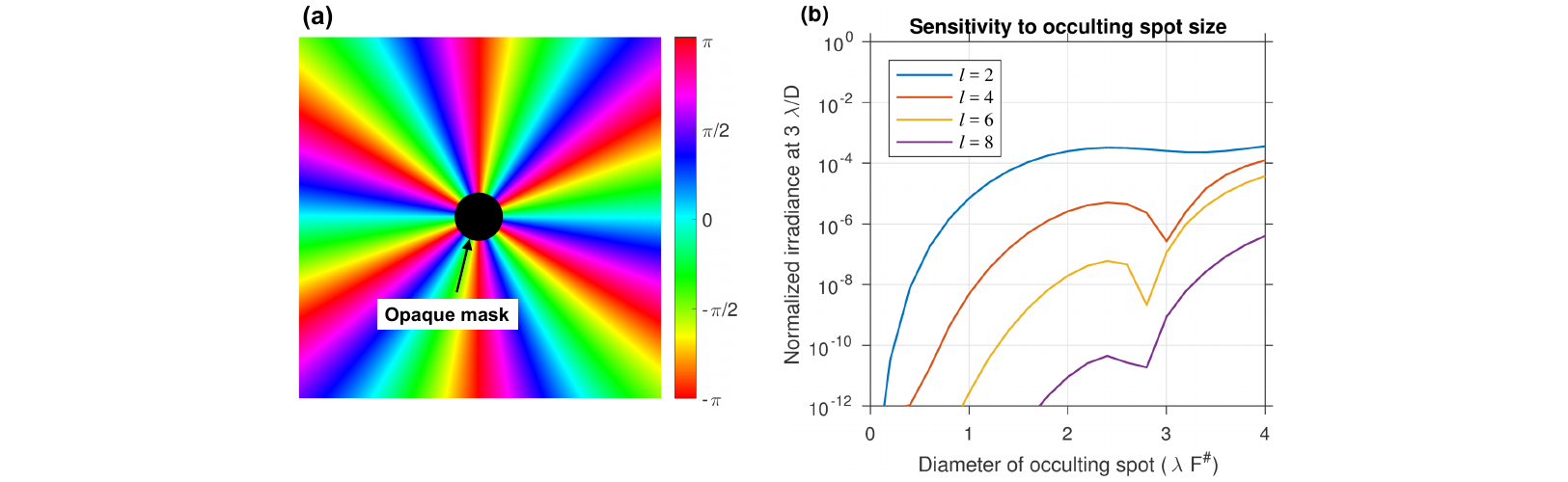}
    \caption{The influence of an opaque spot at the center of the focal plane mask. (a) Phase shift imposed by a charge 6 vortex mask. Light is blocked within the mask at the center (shown in black). (b) Stellar irradiance, averaged over source positions $3\pm0.5~\lambda/D$ and normalized to the peak of the telescope PSF as a function of the mask diameter. Larger masks may be used with higher charges, which alleviate some manufacturing challenges and allow for a reflective low order wavefront sensor.}
    \label{fig:OccMaskLeak}
\end{figure}

\subsection{Effect of adding an opaque spot to the vortex mask}

Manufacturing processes will limit the minimum size of the central defect in a vortex phase mask. A small opaque occulting spot may be introduced to block the central region where the phase shift deviates from the ideal vortex pattern (see Fig. \ref{fig:OccMaskLeak}a). The maximum allowable size of this mask depends on the charge of the vortex. Figure \ref{fig:OccMaskLeak}b shows the stellar irradiance at $3\pm0.5~\lambda/D$ as a function of the mask diameter for various vortex charges. For charge 6 and 8, the occulting mask can be as large as $\sim1~\lambda F^\#$ and $\sim1.7~\lambda F^\#$, respectively, while maintaining sufficient suppression for imaging of Earth-like planets. In each case, the opaque mask does not significantly degrade the planet throughput.

In addition to masking manufacturing errors, there are other potential benefits to introducing a central opaque spot. For instance, the reflection from the spot may be used for integrated low-order wavefront sensing, as recently demonstrated for the WFIRST coronagraph instrument\cite{Shi2017}, potentially in addition to a reflective Lyot stop sensor\cite{Singh2014,Singh2015}. In the case of a charge 6 vortex coronagraph, $\sim$80\% of the starlight would be available from the reflection off of the opaque mask for fast tip-tilt and low-order wavefront sensing. Combined with the natural insensitivity to low order aberrations of vortex coronagraphs, this capability will help maintain deep starlight suppression throughout observations and extend the time between calibrations of the wavefront error and reference star images, thereby improving overall observing efficiency. 

\section{Architecture B: 6.5~m off-axis, unobscured, segmented telescope} \label{sec:offaxis_seg} 

\begin{figure}[t]
    \centering
    \includegraphics[width=\linewidth]{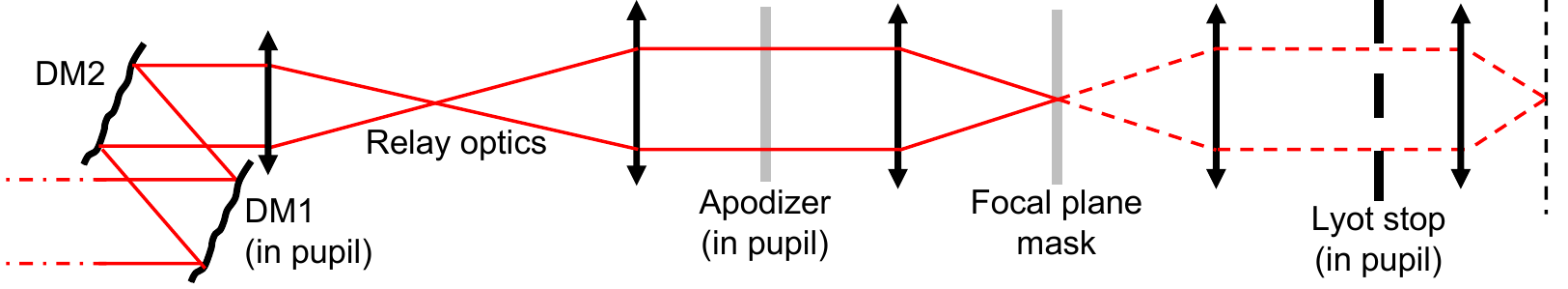}
    \caption{Schematic of an apodized vortex coronagraph. A gray-scale apodizer (see Fig. \ref{fig:apodizedVC_unobs}) prevents unwanted diffraction from the non-circular outer edge of the primary and gaps between mirror segments.}
    \label{fig:adodized_diagram}
\end{figure}

The second potential telescope architecture we study for the HabEx mission concept is a 6.5~m off-axis segmented telescope. This arrangement introduces a few additional complications with respect to the monolithic version. First, a primary mirror with a non-circular outer edge generates diffraction patterns that are difficult to null. To remedy this, we insert a circular sub-aperture in a pupil plane just before the focal plane mask, which provides improved starlight suppression at the cost of throughput (see Fig. \ref{fig:adodized_diagram}). Partial segments may also be introduced to form a circular outer edge. Second, the gaps between mirror segments must be apodized to prevent unwanted diffraction in the image plane from amplitude discontinuities. In this section, we present a promising vortex coronagraph design for the 6.5~m HabEx concept and address the associated telescope requirements. 

\begin{figure}[t]
    \centering
    \includegraphics[width=\linewidth]{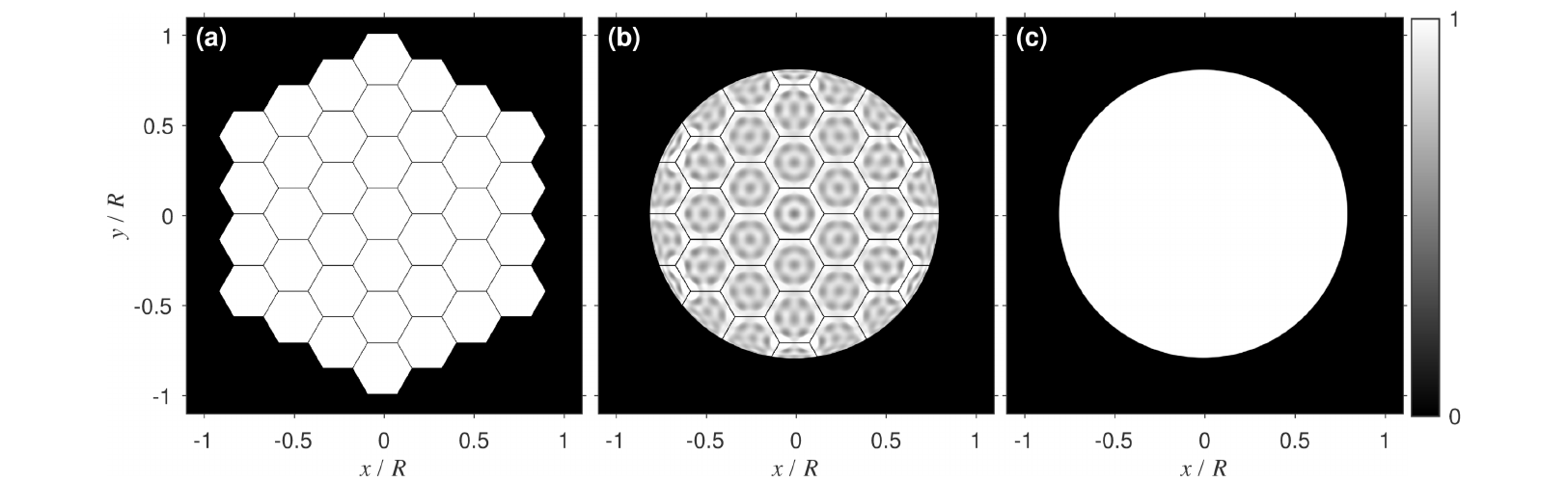}
    \caption{An apodized vortex coronagraph for a 6.5~m HabEx. (a) The image of the primary mirror at the entrance pupil of the coronagraph. (b) The apodizer (squared-magnitude of the desired pupil field). (c) The Lyot stop. The apodizer and Lyot stop diameters are 83\% and 80\% of the pupil diameter (flat-to-flat).}
    \label{fig:apodizedVC_unobs}
\end{figure}

\subsection{Apodized vortex coronagraph design}

Figure \ref{fig:apodizedVC_unobs}a shows a notional primary mirror with 37 hexagonal segments whose widths are $\sim$0.9~m flat-to-flat. The corresponding pupil masks used in the apodized vortex coronagraph are shown in Fig. \ref{fig:apodizedVC_unobs}b,c. The apodizer clips the outer edge of the pupil to make it circular and imparts an amplitude-only apodization pattern on the transmitted or reflected field. Most of the starlight is then diffracted by the vortex outside of the Lyot stop. The small amount of starlight that leaks through the Lyot stop ($\sim$2\%) only contains high-spatial frequencies greater than a specified value $\xi_\text{max}=20$ cycles across the pupil diameter. Thus, in an otherwise perfect optical system, a dark hole appears in the starlight within a 20~$\lambda/D$ radius of the star position for all even nonzero values of the vortex charge $l$. We used the Auxiliary Field Optimization (AFO) method\cite{Jewell2017} to calculate the optimal grayscale pattern\cite{Ruane2016_SPIE,Ruane2017_SPIE}.

\begin{figure}[t]
    \centering
    \includegraphics[width=\linewidth]{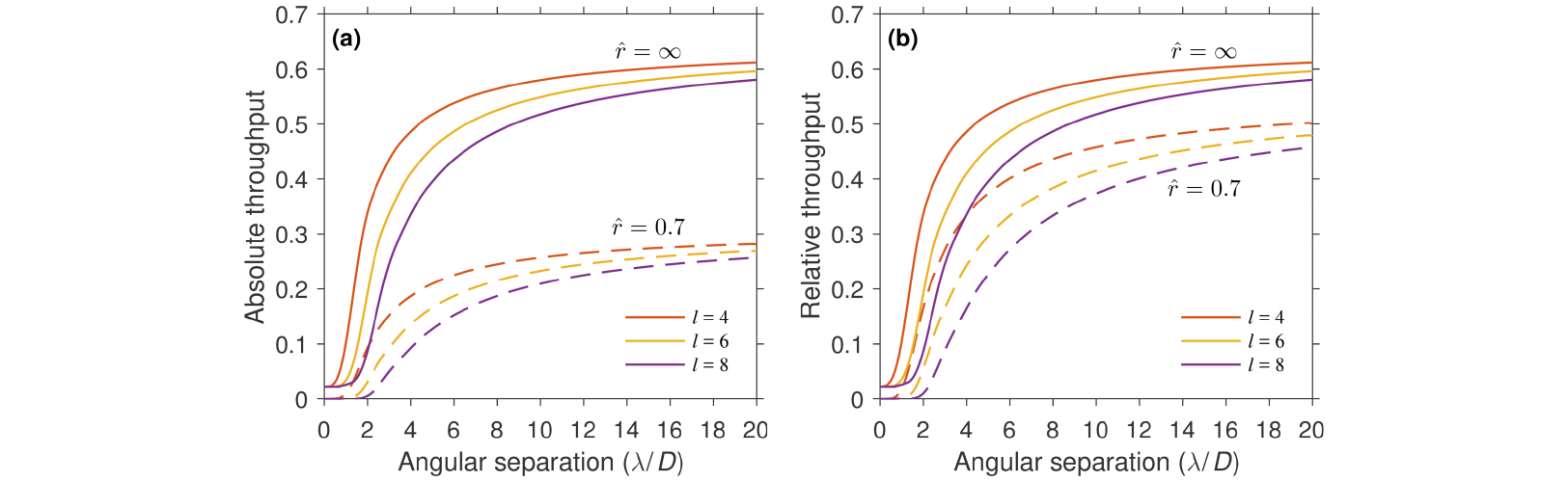}
    \caption{The throughput of the apodized vortex coronagraph with charge 4, 6, and 8 focal plane masks. (a) Absolute throughput. The fraction of total planet light that falls within $\hat{r}\lambda/D$ of the planet position, assuming an otherwise perfect optical system. (b) Relative throughput. The fraction of planet light that falls within $\hat{r}\lambda/D$ of the planet position compared to case with the coronagraph masks removed. Throughput losses originate from introducing a semi-transparent mask and clipping the outer edge of the pupil to create a circular boundary (see Fig. \ref{fig:apodizedVC_unobs}).}
    \label{fig:Thpt_apodVC_unobs}
\end{figure}

The throughput of the coronagraph with various focal plane vortex masks is shown in Figure \ref{fig:Thpt_apodVC_unobs}. We report both the absolute throughput $\eta_p$ and relative throughput $\eta_p/\eta_\text{tel}$ within a circular region of interest of radius $\hat{r}\lambda/D$ centered on the planet position, where $\eta_\text{tel}$ represents the throughput of the telescope with the coronagraph masks removed. After the coronagraph, $\sim60\%$ of the total energy from an off-axis source remains. Less than 30$\%$ of the total energy appears within $0.7~\lambda/D$ of the planet position, including losses from the apodizer and broadening of the point spread function by the undersized pupil mask and Lyot stop. Approximately 50\% of the planet light remains within $0.7~\lambda/D$ compared to the point spread function with the coronagraph masks removed. Other than a loss in throughput, the apodized version shares most of the same performance characteristics as the conventional vortex coronagraph. Furthermore, a primary mirror that includes partial segments to create a circular outer boundary would allow for drastically improved coronagraph throughput. 

\begin{figure}[t!]
    \centering
    \includegraphics[width=0.85\linewidth]{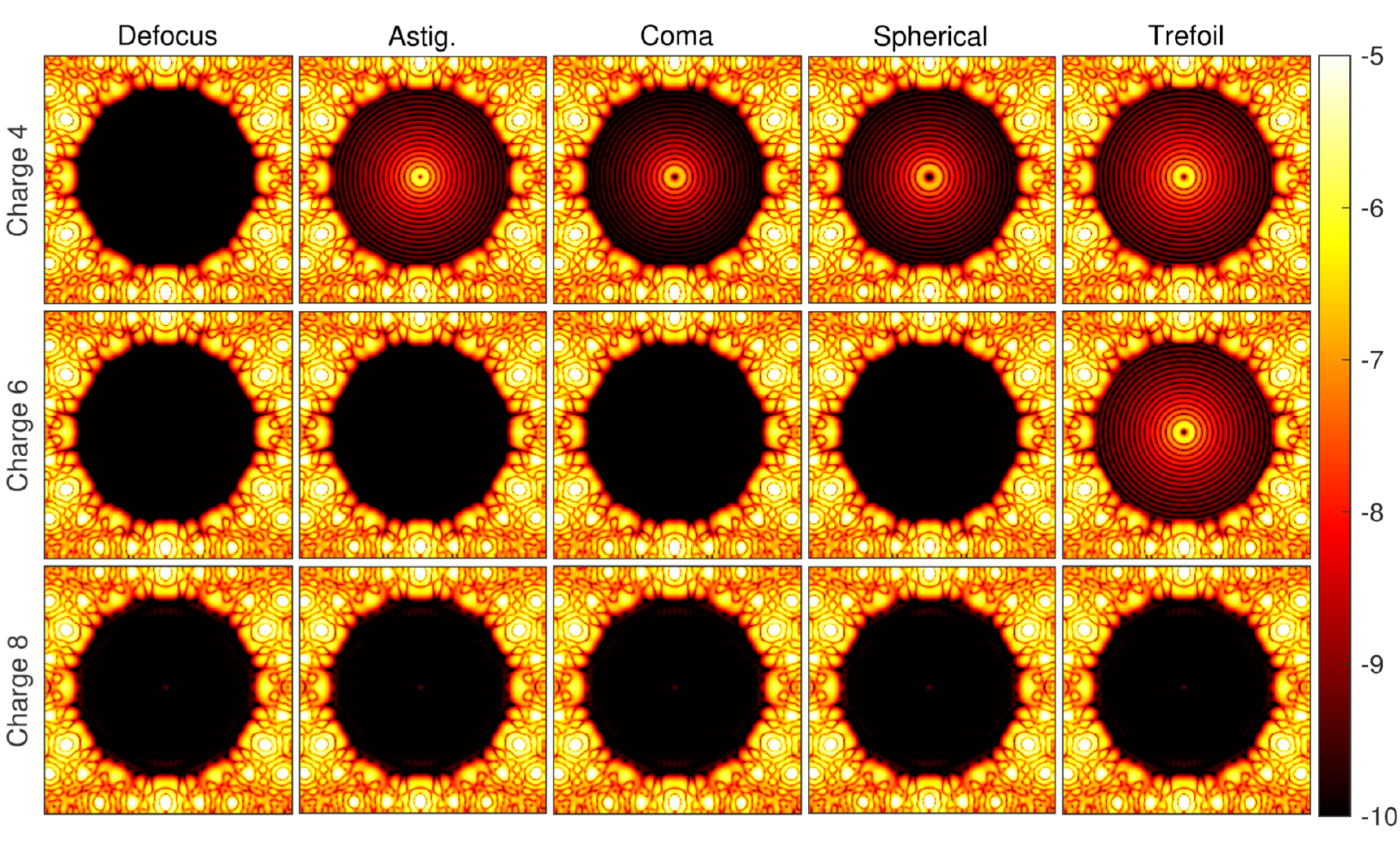}
    \caption{The sensitivity of an apodized vortex coronagraph to low order aberrations on an off-axis, segmented telescope. Log irradiance owing to $\lambda/1000$ rms wavefront error in each mode, normalized to the peak value with the coronagraph masks removed. The dark zone has an angular diameter of 40$\lambda/D$. As in the case of a monolithic telescope, higher charge vortex coronagraphs passively suppress more low order Zernike modes.}
    \label{fig:apodizedVC_unobs_Zsens}
\end{figure}
\begin{figure}[t!]
    \centering
    \includegraphics[width=0.85\linewidth]{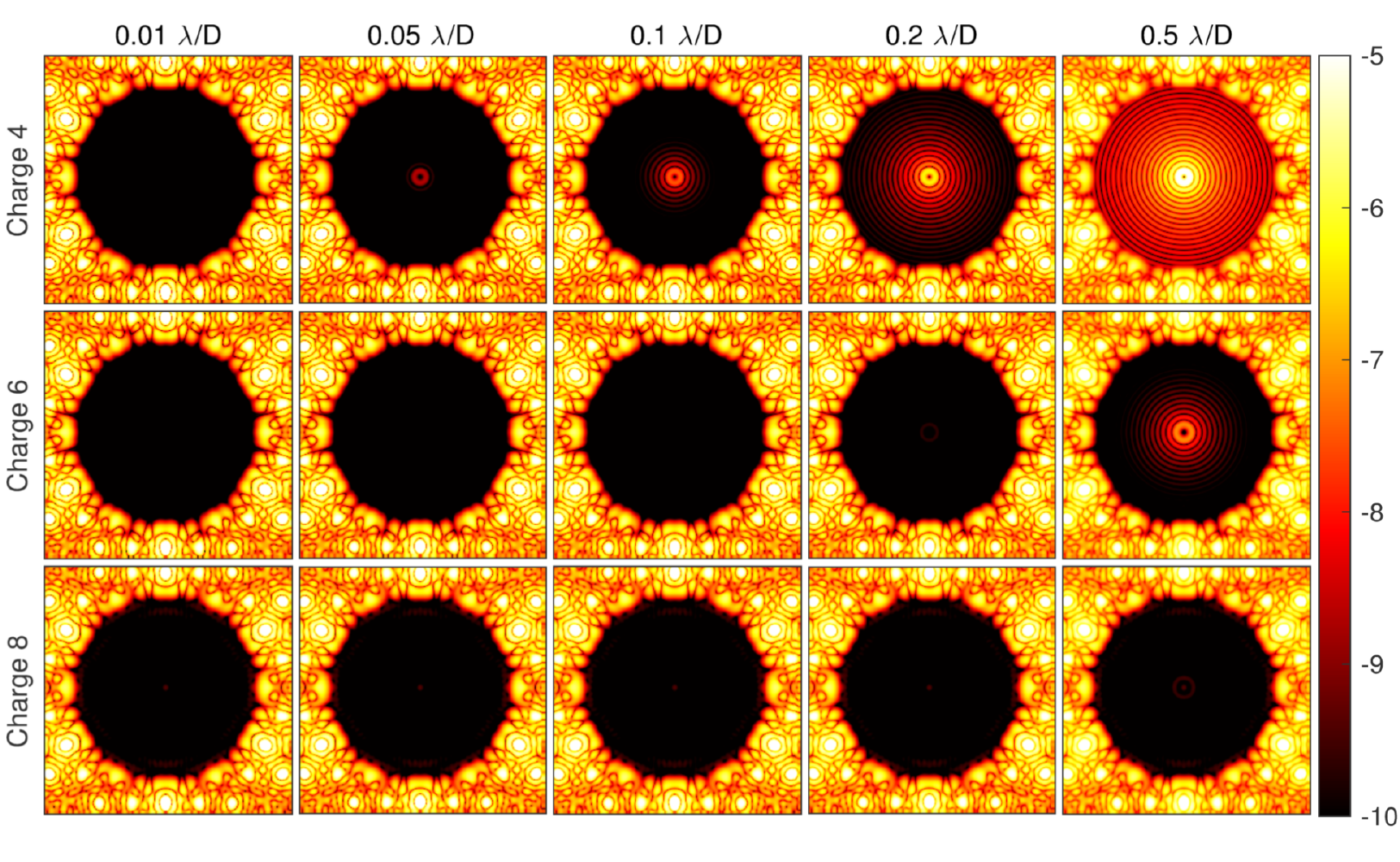}
    \caption{The sensitivity of an apodized vortex coronagraph to stellar angular diameter on an off-axis, segmented telescope. Log stellar irradiance, normalized to the peak value with the coronagraph masks removed. The dark zone has a diameter of 40$\lambda/D$. The simulation is monochromatic, but applies to all wavelengths. As in the case of a monolithic telescope, higher charge vortex coronagraphs are leak less light from partially resolved stars.}
    \label{fig:apodizedVC_unobs_FiniteStarSens}
\end{figure}

\subsection{Sensitivity to low-order aberrations and the angular size of stars}

Assuming the telescope is off-axis and unobstructed, the leaked stellar irradiance in the presence of low-order aberrations appears identical to the monolithic case, up to a radius of $\xi_\text{max} \lambda F^{\#}$ (see Fig. \ref{fig:apodizedVC_unobs_Zsens}). However, to maintain a fixed raw contrast threshold, the wavefront error requirements presented in Table \ref{tab:loworder} scale as $1/\sqrt{\eta_p}$; i.e. get tougher to guarantee than in the monolithic case. For the sake of brevity, we have not included an updated wavefront error requirement table here. 

The stellar leakage due to the angular size of stars is also equivalent to a vortex coronagraph without an apodizer. Figure \ref{fig:apodizedVC_unobs_FiniteStarSens} shows the leaked starlight as a function of stellar angular size. A charge 4 is sufficient to suppress stars $\lesssim0.01~\lambda/D$ in diameter. A charge 6 or 8 may be used to maximize SNR ($\eta_p/\sqrt{\eta_s})$ in the case of a larger star, such as Alpha Centauri A whose angular diameter is 8.5~mas or $\sim$0.5~$\lambda/D$ in the visible.

\begin{figure}
    \centering
    \includegraphics[width=\linewidth]{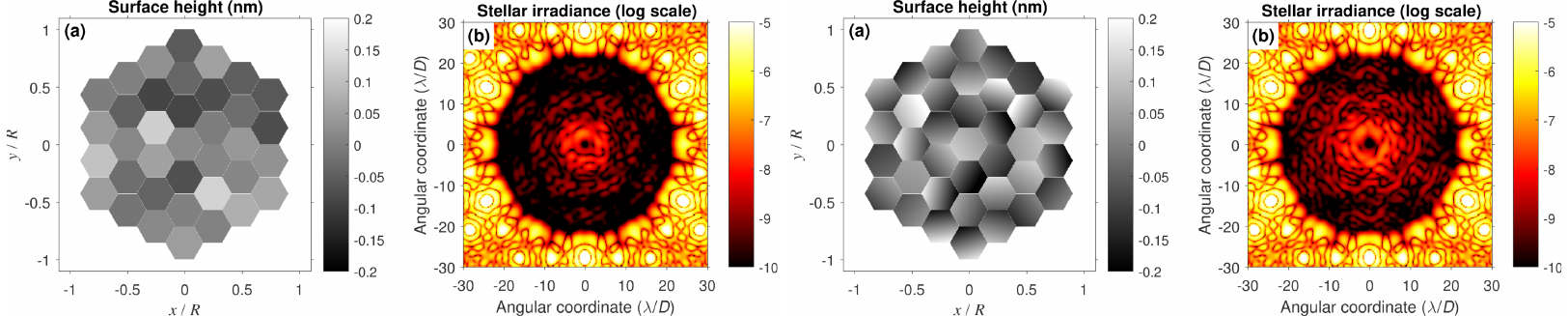}

    \caption{\textit{left:} (a) Example wavefront with 100~pm rms of random segment piston errors and (b) the corresponding stellar irradiance at $\lambda=450~\text{nm}$. \textit{right:} Same as \textit{left}, but with an additional 0.005~$\lambda/D=71~\mu$as~rms of random tip-tilt errors. These aberrations cause speckles to appear close to the star where planets are likely to reside.}
    \label{fig:segErrs}
\end{figure}

\begin{figure}
    \centering
    \includegraphics[width=\linewidth]{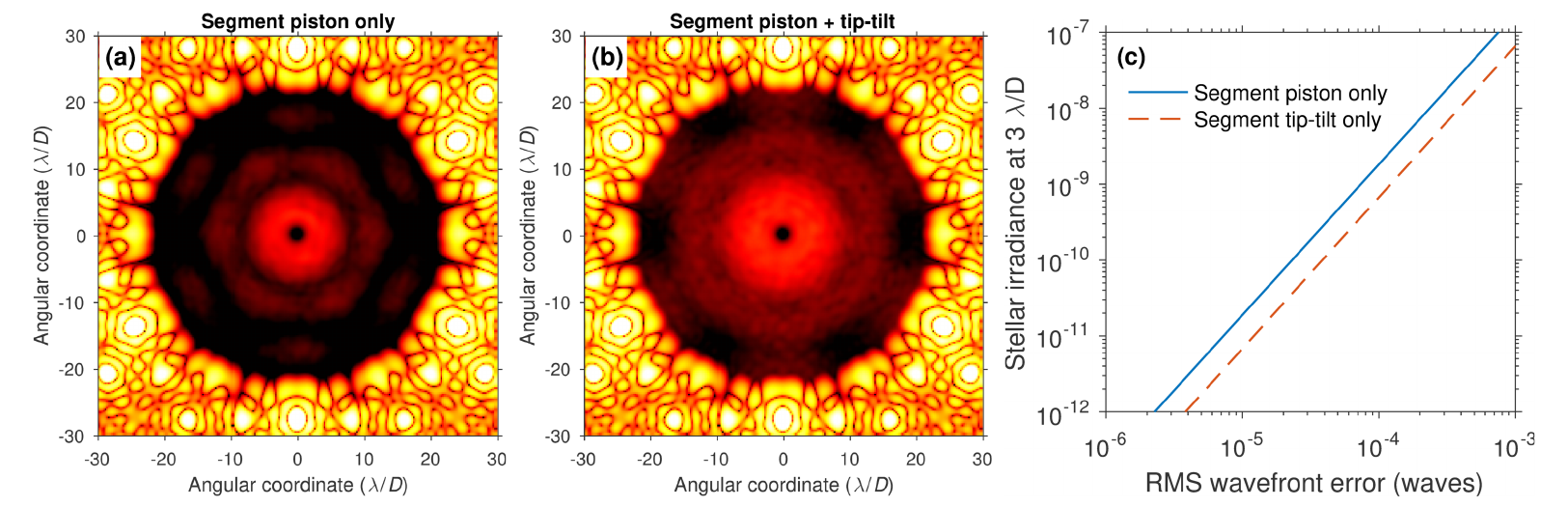}
    \caption{(a)-(b) Time average over many realizations of leaked stellar irradiance at $\lambda=450$~nm due to (a) 100~pm rms of random segment piston and (b) with an additional 0.005~$\lambda/D=71~\mu$as~rms tip-tilt error. (c) Dependence of stellar irradiance at 3~$\lambda/D=43$~mas on the rms wavefront error segment piston and tip-tilt. The simulation is monochromatic, but applies to all wavelengths. Segment piston and tip-tilt must be controlled to $10^{-5}$~waves~rms, or $<$10~pm~rms, to ensure detection of Earth-like planets.}
    \label{fig:segErrs2}
\end{figure}

\subsection{Segment co-phasing requirements}

A major challenge for exoplanet imaging with a segmented telescope will be to keep the mirrors co-aligned throughout observations. As shown in Fig.~\ref{fig:segErrs}, small segment motions in piston and tip-tilt cause speckles to appear in the dark hole which may be difficult to calibrate and will likely contribute to both photon and spatial speckle noise. Figure~\ref{fig:segErrs2}a-b shows the time average over many realization of the errors shown in Fig. \ref{fig:segErrs} drawn from Gaussian distributions for both piston and tip-tilt with standard deviations of 100 pm and 0.005~$\lambda/D=71~\mu$as~rms. When the mirror segments have random piston errors only, the resulting distribution of light resembles the diffraction pattern of a single segment (Fig.~\ref{fig:segErrs2}a). Random segment tip-tilt error tend to spread the leaked starlight to larger separations (Fig.~\ref{fig:segErrs2}b). However, the leaked starlight is well approximated by a similar second order power law in both cases (Fig.~\ref{fig:segErrs2}c), which yields a wavefront error requirement of $\sim$10~pm~rms, similar in magnitude to unsuppressed low-order modes. On the other hand, if the primary mirror segments undergo a coordinated movement that resembles a low-order Zernike polynomial $Z_n^m$, the amount of leaked starlight would be significantly smaller if $l>n+|m|$ and the tolerance to such a motion would be considerably relaxed.

\subsection{Fabrication of grayscale apodizing masks}

Achromatic grayscale apodizers have been fabricated using metallic microdots arranged in error-diffused patterns \cite{Dorrer2007,Sivaramakrishnan2009}. Prototype masks produced specifically for the purpose of demonstrating apodized vortex coronagraphs are currently being tested at the High Contrast Spectroscopy Testbed for Segmented Telescopes (HCST) at Caltech \cite{Delorme2017}. These experiments seek to validate this approach for use on HabEx and prepare for future testing on vacuum testbeds.

\section{Conclusions and outlook} \label{sec:conc} 

Vortex coronagraphs provide a viable pathway towards imaging Earth-like exoplanets with the HabEx decadal mission concept with a fully off-axis telescope architecture. We have provided an overview of the performance of vortex coronagraphs and wavefront stability requirements. The off-axis design of the HabEx telescope allows for the best possible performance in terms of throughput, inner working angle, and robustness to aberrations. With a vortex coronagraph, the low order wavefront error requirements for imaging Earth-like planets with HabEx are comparable to those to be demonstrated by the WFIRST-CGI for imaging Jupiters \cite{Krist2015}. A segmented primary mirror does not fundamentally change the performance characteristics of a vortex coronagraph. However, mirror segments introduce challenging segment co-phasing requirements and the need for apodization. In addition to the grayscale pupil mask presented here, alternate apodization approaches are available that shape the pupil amplitude using deformable mirrors \cite{Mazoyer2017}. A trade study is needed to identify the performance trades between these apodization solutions. In all cases, we find that the throughput of vortex coronagraphs and robust to wavefront errors degrades significantly on centrally-obscured telescopes. Several studies are underway to improve performance for such architectures \cite{Fogarty2017}.

\appendix

\section{Zernike polynomials}

The Zernike polynomials \cite{Zernike1934} may be written as 
\begin{equation}
    Z_n^m\left(r/a,\theta\right) = R_n^{|m|}\left(r/a\right)
    \begin{cases} 
        \cos\left(m\theta\right) & m \geq 0 \\
        \sin\left(|m|\theta\right) & m<0
    \end{cases}
    ,\;\;\;\;\;r\le a,
\label{eq:Zpupil_appendix}
\end{equation}
where $R_n^m(r/a)$ is the radial Zernike polynomial given by 
\begin{equation}
R_n^m(r/a) = \sum_{k=0}^{\frac{n-m}{2}}\frac{(-1)^k(n-k)!}{k!\left(\frac{n+m}{2}-k\right)!\left(\frac{n-m}{2}-k\right)!}(r/a)^{n-2k},\;\;\;\;\;r/a\le 1,
\end{equation}
where $n-m$ is even. The indices $n$ and $m$ are integers respectively known as the degree and azimuthal order. The first few radial polynomials are: $R_0^0=1$, $R_1^1=r/a$, $R_2^0=2(r/a)^2-1$, $R_2^2=(r/a)^2$, $R_3^1=3(r/a)^3-2(r/a)$, $R_3^3=(r/a)^3$. 

\section{Weber-Schafheitlin integrals}

The pupil functions generated by vortex coronagraphs are a subset of solutions of the discontinuous integral of Weber and Schafheitlin \cite{Watson1922}, which in its conventional form is written
\begin{equation}
W_{\nu,\mu,\lambda}(t;\alpha,\beta)=\int_0^\infty{ \frac{J_{\nu}(\alpha t)J_{\mu}(\beta t)}{t^\lambda} dt},
\label{eqn:WSintegral}
\end{equation}
where $\nu,\mu,\lambda$ are integers and $\alpha$ and $\beta$ are constants. The integral is convergent provided $\nu+\mu-\lambda \ge 0$ and $\lambda \ge 0$. If $0<\alpha<\beta$,
\begin{align}
W_{\nu,\mu,\lambda}(t;\alpha,\beta)=&\frac{\alpha^{\nu}\Gamma \left( \frac{\nu + \mu - \lambda + 1}{2} \right)}{2^{\lambda}\beta^{\nu-\lambda+1}\Gamma \left( \frac{-\nu + \mu + \lambda + 1}{2} \right) \Gamma(\nu+1)} \\ 
&\times \prescript{}{2}F_{1}\left( \frac{\nu + \mu - \lambda + 1}{2},\frac{\nu - \mu - \lambda + 1}{2};\nu+1;\frac{\alpha^2}{\beta^2} \right),
\end{align}
where $\Gamma(~)$ is the gamma function and $\prescript{}{2}F_{1}(~)$ is a hypergeometric function \cite{Gradshteyn}. On the other hand, if $0<\beta<\alpha$
\begin{align}
W_{\nu,\mu,\lambda}(t;\alpha,\beta)=&\frac{\beta^{\nu}\Gamma \left( \frac{\nu + \mu - \lambda + 1}{2} \right)}{2^{\lambda}\alpha^{\nu-\lambda+1}\Gamma \left( \frac{\nu - \mu + \lambda + 1}{2} \right) \Gamma(\nu+1)} \\ &\times \prescript{}{2}F_{1}\left( \frac{\nu + \mu - \lambda + 1}{2},\frac{-\nu + \mu - \lambda + 1}{2};\mu+1;\frac{\beta^2}{\alpha^2} \right).
\end{align}
Integrals with the form of Eqn. \ref{eqn:WSintegral}, namely a product of Bessel functions, appear in the output function integral in cases where the input function is circular or may be described by a Zernike polynomial in amplitude \cite{Ruane2015_SPIE}.

\section{First order exit pupil modes}
Here, we provide the analytical solutions to Eq. \ref{eqn:glnm} for $l\ge 0$ and $m \ge 0$: 

\subsection{Piston $Z_0^0$}
\begin{align}
    g_{000}(r,\theta)=&
    \begin{cases}
        1 & r \le a \\
        0 & r > a
    \end{cases}\\
    g_{002}(r,\theta)=&
    \begin{cases}
        0 & r \le a \\
        \left(\frac{a}{r} \right)^2e^{i2\theta} & r > a
    \end{cases}\\
    g_{004}(r,\theta)=&
    \begin{cases}
        0 & r \le a \\
        \left[3\left(\frac{a}{r} \right)^4+2\left(\frac{a}{r} \right)^2\right]e^{i4\theta} & r > a
    \end{cases}\\
    g_{006}(r,\theta)=&
    \begin{cases}
        0 & r \le a \\
        \left[10\left(\frac{a}{r} \right)^6-12\left(\frac{a}{r} \right)^4+3\left(\frac{a}{r} \right)^2\right]e^{i6\theta} & r > a
    \end{cases}\\
    g_{008}(r,\theta)=&
    \begin{cases}
        0 & r \le a \\
        \left[35\left(\frac{a}{r} \right)^8-60\left(\frac{a}{r} \right)^6+30\left(\frac{a}{r} \right)^4-4\left(\frac{a}{r} \right)^2\right]e^{i8\theta} & r > a
    \end{cases}
\end{align}

\subsection{Tip-tilt $Z_1^1$}
\begin{align}
    g_{110}(r,\theta)=&
    \begin{cases}
        \frac{r}{a}\cos(\theta) & r \le a \\
        0 & r > a
    \end{cases}\\
    g_{112}(r,\theta)=&\frac{1}{2}
    \begin{cases}
        \frac{r}{a}e^{i\theta} & r \le a \\
        \left(\frac{a}{r}\right)^3e^{i3\theta} & r > a
    \end{cases}\\
g_{114}(r,\theta)=&\frac{1}{2}
    \begin{cases}
        0 & r \le a \\
        \left[4\left(\frac{a}{r}\right)^5 - 3\left(\frac{a}{r}\right)^3\right]e^{i5\theta} + \left(\frac{a}{r}\right)^3e^{i3\theta} & r > a
    \end{cases}\\
g_{116}(r,\theta)=&\frac{1}{2}
    \begin{cases}
        0 & r \le a \\
        \left[15\left(\frac{a}{r}\right)^7 - 20\left(\frac{a}{r}\right)^5 + 6 \left(\frac{a}{r}\right)^3 \right]e^{i7\theta} + \left[4\left(\frac{a}{r}\right)^5 - 3\left(\frac{a}{r}\right)^3\right]e^{i5\theta} & r > a
    \end{cases}\\
g_{118}(r,\theta)=&\frac{1}{2}
    \begin{cases}
        0 & r \le a \\
        \left[56\left(\frac{a}{r}\right)^9 -  105\left(\frac{a}{r}\right)^7 + 60\left(\frac{a}{r}\right)^5 - 10 \left(\frac{a}{r}\right)^3 \right]e^{i9\theta} + \left[15\left(\frac{a}{r}\right)^7 - 20\left(\frac{a}{r}\right)^5 + 6 \left(\frac{a}{r}\right)^3 \right]e^{i7\theta} & r > a
    \end{cases}
\end{align}

\subsection{Defocus $Z_2^0$}
\begin{align}
    g_{200}(r,\theta)=&
    \begin{cases}
        2\left(\frac{r}{a}\right)^2-1 & r \le a \\
        0 & r > a
    \end{cases}\\
    g_{202}(r,\theta)=&
    \begin{cases}
        \left(\frac{r}{a}\right)^2 e^{i2\theta} & r \le a \\
        0 & r > a
    \end{cases}\\
    g_{204}(r,\theta)=&
    \begin{cases}
        0 & r \le a \\
        \left(\frac{a}{r}\right)^4 e^{i4\theta} & r > a
    \end{cases}\\
    g_{206}(r,\theta)=&
    \begin{cases}
        0 & r \le a \\
        \left[5\left(\frac{a}{r}\right)^6-4\left(\frac{a}{r}\right)^4\right] e^{i6\theta} & r > a
    \end{cases}\\
    g_{208}(r,\theta)=&
    \begin{cases}
        0 & r \le a \\
        \left[21\left(\frac{a}{r}\right)^8-30\left(\frac{a}{r}\right)^6+10\left(\frac{a}{r}\right)^4\right] e^{i8\theta} & r > a
    \end{cases}
\end{align}

\subsection{Astigmatism $Z_2^2$}
\begin{align}
    g_{220}(r,\theta)=&
    \begin{cases}
        \left(\frac{r}{a}\right)^2\cos(2\theta) & r \le a \\
        0 & r > a
    \end{cases}\\
    g_{222}(r,\theta)=&\frac{1}{2}
    \begin{cases}
        2\left(\frac{r}{a}\right)^2-1 & r \le a \\
        \left(\frac{a}{r}\right)^4e^{i4\theta} & r > a
    \end{cases}\\
    g_{224}(r,\theta)=&\frac{1}{2}
    \begin{cases}
        \left(\frac{r}{a}\right)^2e^{i2\theta} & r \le a \\
        \left[5\left(\frac{a}{r}\right)^6-4\left(\frac{a}{r}\right)^4\right]e^{i6\theta} & r > a
    \end{cases}\\
    g_{226}(r,\theta)=&
    \begin{cases}
        0 & r \le a \\
        \left[\frac{21}{2}\left(\frac{a}{r}\right)^8-15\left(\frac{a}{r}\right)^6+5\left(\frac{a}{r}\right)^4\right]e^{i8\theta} + \frac{1}{2}\left(\frac{a}{r}\right)^4e^{i4\theta} & r > a
    \end{cases}\\
    g_{228}(r,\theta)=&
    \begin{cases}
        0 & r \le a \\
        \left[42\left(\frac{a}{r}\right)^{10}-84\left(\frac{a}{r}\right)^8+\frac{105}{2}\left(\frac{a}{r}\right)^6-10\left(\frac{a}{r}\right)^4\right]e^{i10\theta} + \left[\frac{5}{2}\left(\frac{a}{r}\right)^6-2\left(\frac{a}{r}\right)^4\right]e^{i6\theta} & r > a
    \end{cases}
\end{align}

\subsection{Coma $Z_3^1$}
\begin{align}
    g_{310}(r,\theta)=&
    \begin{cases}
        \left[3\left(\frac{r}{a}\right)^3-2\frac{r}{a}\right]\cos(\theta) & r \le a \\
        0 & r > a
    \end{cases}\\
    g_{312}(r,\theta)=&
    \begin{cases}
        \frac{1}{2}\left( \frac{r}{a} \right)^3e^{i3\theta} + \left[ \frac{3}{2}\left( \frac{r}{a} \right)^3 - \frac{r}{a} \right]e^{i\theta} & r \le a \\
        0 & r > a
    \end{cases}\\
    g_{314}(r,\theta)=&\frac{1}{2}
    \begin{cases}
        \left(\frac{r}{a}\right)^3e^{i3\theta} & r \le a \\
        \left(\frac{a}{r}\right)^5e^{i5\theta}  & r > a
    \end{cases}\\
    g_{316}(r,\theta)=&
    \begin{cases}
        0 & r \le a \\
        \left[3\left(\frac{a}{r}\right)^7-\frac{5}{2}\left(\frac{a}{r}\right)^5\right]e^{i7\theta} + \frac{1}{2}\left(\frac{a}{r}\right)^5e^{i5\theta} & r > a
    \end{cases}\\
    g_{318}(r,\theta)=&
    \begin{cases}
        0 & r \le a \\
        \left[14\left(\frac{a}{r}\right)^{9}-21\left(\frac{a}{r}\right)^7+\frac{15}{2}\left(\frac{a}{r}\right)^5\right]e^{i9\theta} + \left[3\left(\frac{a}{r}\right)^7-\frac{5}{2}\left(\frac{a}{r}\right)^5\right]e^{i7\theta} & r > a
    \end{cases}
\end{align}

\subsection{Spherical $Z_4^0$}
\begin{align}
    g_{400}(r,\theta)=&
    \begin{cases}
        6\left(\frac{r}{a}\right)^4-6\left(\frac{r}{a}\right)^2+1 & r \le a \\
        0 & r > a
    \end{cases}\\
    g_{402}(r,\theta)=&
    \begin{cases}
        \left[4\left(\frac{r}{a}\right)^4-3\left(\frac{r}{a}\right)^2\right] e^{i2\theta} & r \le a \\
        0 & r > a
    \end{cases}\\
    g_{404}(r,\theta)=&
    \begin{cases}
        \left(\frac{r}{a}\right)^4 e^{i4\theta} & r \le a \\
        0 & r > a
    \end{cases}\\
    g_{406}(r,\theta)=&
    \begin{cases}
        0 & r \le a \\
        \left(\frac{a}{r}\right)^6 e^{i6\theta} & r > a
    \end{cases}\\
    g_{408}(r,\theta)=&
    \begin{cases}
        0 & r \le a \\
        \left[7\left(\frac{a}{r}\right)^8-6\left(\frac{a}{r}\right)^6\right] e^{i8\theta} & r > a
    \end{cases}
\end{align}

\subsection{Trefoil $Z_3^3$}

\begin{align}
    g_{330}(r,\theta)=&
    \begin{cases}
        \left(\frac{r}{a}\right)^3\cos(3\theta) & r \le a \\
        0 & r > a
    \end{cases}\\
    g_{332}(r,\theta)=&\frac{1}{2}
    \begin{cases}
        \left[3\left(\frac{r}{a}\right)^3-2\frac{r}{a}\right] e^{-i\theta} & r \le a \\
        \left(\frac{a}{r}\right)^5 e^{i5\theta} & r > a
    \end{cases}\\
    g_{334}(r,\theta)=&\frac{1}{2}
    \begin{cases}
        \left[3\left(\frac{r}{a}\right)^3-2\frac{r}{a}\right] e^{i\theta} & r \le a \\
        \left[6\left(\frac{a}{r}\right)^3-5\left(\frac{a}{r}\right)^5\right] e^{i7\theta} & r > a
    \end{cases}\\
    g_{336}(r,\theta)=&\frac{1}{2}
    \begin{cases}
        \left(\frac{r}{a}\right)^3 e^{i3\theta} & r \le a \\
        \left[28\left(\frac{a}{r}\right)^9-42\left(\frac{a}{r}\right)^7+15\left(\frac{a}{r}\right)^5\right] e^{i9\theta} & r > a
    \end{cases}\\
    g_{338}(r,\theta)=&
    \begin{cases}
        0 & r \le a \\
        \left[60\left(\frac{a}{r}\right)^{11}-126\left(\frac{a}{r}\right)^9+84\left(\frac{a}{r}\right)^7-\frac{35}{2}\left(\frac{a}{r}\right)^5 \right] e^{i11\theta} + \frac{1}{2}\left(\frac{a}{r}\right)^5 e^{i5\theta} & r > a
    \end{cases}
\end{align}

\clearpage
\section{Tables}

\begin{table}[h!]
\caption{Low-order wavefront error requirements for Earth-like exoplanet detection with vortex coronagraphs on future off-axis, monolithic, space telescopes.}
\label{tab:loworder}
\begin{center}       
\begin{tabular}{|c|P{0.95cm}|P{0.95cm}|P{0.95cm}|P{1.7cm}|P{1.7cm}|P{1.7cm}|P{1.7cm}|}
\hline
\rule[-1ex]{0pt}{3.5ex} Aberration & \multicolumn{3}{c}{Indices} & \multicolumn{4}{|c|}{Allowable RMS wavefront error per mode (nm) }\\
\hline
\rule[-1ex]{0pt}{3.5ex}  & Noll & $n$ & $m$ & $l=4$ & $l=6$ & $l=8$ & $l=10$\\
\hline\hline
\rule[-1ex]{0pt}{3.5ex} Tip-tilt & 2,3 & 1 & $\pm$1 & 1.1 & 5.9 & 14 & 26 \\
\hline
\rule[-1ex]{0pt}{3.5ex} Defocus & 4 & 2 & 0 & 0.81 & 4.6 & 12 & 26 \\
\hline
\rule[-1ex]{0pt}{3.5ex} Astigmatism  & 5,6 & 2 & $\pm$2 & 0.007 & 1.1 & 0.9 & 4.6 \\
\hline
\rule[-1ex]{0pt}{3.5ex} Coma & 7,8 & 3 & $\pm$1 & 0.006 & 0.66 & 0.82 & 5 \\
\hline
\rule[-1ex]{0pt}{3.5ex} Trefoil & 9,10 & 3 & $\pm$3 & 0.007 & 0.006 & 0.57 & 0.67 \\
\hline
\rule[-1ex]{0pt}{3.5ex} Spherical & 11 & 4 & 0 & 0.005 & 0.51 & 0.73 & 6.3 \\
\hline
\rule[-1ex]{0pt}{3.5ex} $2^\text{nd}$ Astig. & 12,13 & 4 & $\pm$2 & 0.008 & 0.007 & 0.67 & 0.73 \\
\hline
\rule[-1ex]{0pt}{3.5ex} Quadrafoil & 14,15 & 4 & $\pm$4 & 0.008 & 0.008 & 0.006 & 0.54 \\
\hline
\rule[-1ex]{0pt}{3.5ex} $2^\text{nd}$ Coma & 16,17 & 5 & $\pm$1 & 0.004 & 0.005 & 0.69 & 0.85 \\
\hline
\rule[-1ex]{0pt}{3.5ex} $2^\text{nd}$ Trefoil & 18,19 & 5 & $\pm$3 & 0.005 & 0.006 & 0.004 & 0.72 \\
\hline
\rule[-1ex]{0pt}{3.5ex} Pentafoil & 20,21 & 5 & $\pm$5 & 0.005 & 0.005 & 0.005 & 0.005 \\
\hline
\rule[-1ex]{0pt}{3.5ex} $2^\text{nd}$ Spherical & 22 & 6 & 0 & 0.003 & 0.003 & 0.84 & 1.1 \\
\hline
\rule[-1ex]{0pt}{3.5ex} $3^\text{rd}$ Astig. & 23,24 & 6 & $\pm$2 & 0.002 & 0.004 & 0.003 & 0.82 \\
\hline
\rule[-1ex]{0pt}{3.5ex} $2^\text{nd}$ Quadrafoil & 25,26 & 6 & $\pm$4 & 0.003 & 0.003 & 0.003 & 0.004 \\
\hline
\rule[-1ex]{0pt}{3.5ex} Hexafoil & 27,28 & 6 & $\pm$6 & 0.003 & 0.003 & 0.003 & 0.004 \\
\hline
\end{tabular}
\end{center}
\end{table} 

\begin{table}[h!]
\caption{Coefficients for analytical approximations of transmitted energy from point sources at small angular separations, $T_{\alpha}=\tau_l (\pi \alpha D/\lambda)^l$, and extended sources $T_{\Theta}=\kappa_l (\pi \alpha D/\lambda)^l$.}
\label{tab:tiptilt_coeffs}
\begin{center}       
\begin{tabular}{|c|c|c|}
\hline
\rule[-1ex]{0pt}{3.5ex} Charge & $\tau_l$ & $\kappa_l$\\
\hline\hline
\rule[-1ex]{0pt}{3.5ex} $l=2$ & 1/8 & 1/64 \\
\hline
\rule[-1ex]{0pt}{3.5ex} $l=4$ & 1/192 & 1/9216  \\
\hline
\rule[-1ex]{0pt}{3.5ex} $l=6$ & 1/9216 & 1/2359296 \\
\hline
\rule[-1ex]{0pt}{3.5ex} $l=8$ & 1/737280 & 1/943718400  \\
\hline
\end{tabular}
\end{center}
\end{table}

\clearpage

\acknowledgments     
We thank the HabEx Coronagraph Technology Working Group (CTWG) for useful discussions. G. Ruane is supported by an NSF Astronomy and Astrophysics Postdoctoral Fellowship under award AST-1602444. This work was supported by the Exoplanet Exploration Program (ExEP), Jet Propulsion Laboratory, California Institute of Technology, under contract to NASA.


\bibliography{RuaneLibrary}   
\bibliographystyle{spiebib}   

\end{document}